\documentclass[letterpaper, 10 pt, conference]{IEEEtran}

\IEEEoverridecommandlockouts   
\let\proof\relax

\usepackage{hyperref}       
\usepackage{url}            
\usepackage{booktabs}       
\usepackage{amsfonts}       
\usepackage{nicefrac}       
\usepackage{microtype}      
\usepackage{amsmath,amssymb,amsfonts}
\usepackage{algorithmic}
\usepackage{mathrsfs}
\usepackage{graphicx}
\usepackage{textcomp}
\usepackage{xcolor}
\usepackage{cleveref}
\usepackage{dsfont}
\usepackage{amsthm}
\newtheorem{theorem}{Theorem}[]

\newtheorem{lemma}[]{Lemma}

\newtheorem*{assumption*}{Assumption}
\newtheorem{assumption}{Assumption}
\newtheorem{prop}{Proposition}
\newtheorem{definition}{Definition}
\newtheorem*{remark}{Remark}

\DeclareMathOperator*{\argmin}{arg\,min}
\DeclareMathOperator*{\E}{\mathbb{E}}
\DeclareMathOperator*{\R}{\mathbb{R}}

\usepackage{mathtools}
\graphicspath{ {./images/} }
\usepackage{dsfont}
\usepackage[ruled,vlined]{algorithm2e}
\newtheorem{example}{Example}

\date{}

\usepackage[draft]{changes}
	\definechangesauthor[name={Tao Li}, color=orange]{TL}
\begin{document}

\title{On the Resilience of Traffic Networks under Non-Equilibrium Learning\\
}

\author{Yunian Pan, Tao Li, and Quanyan Zhu$^*$
\thanks{$^*$The authors are with the Department of Electrical and Computer Engineering, Tandon School of Engineering, New York University, Brooklyn, NY, 11201 USA; E-mail: {\tt\small \{yp1170,tl2636,qz494\}@nyu.edu}}%
}

\maketitle

\begin{abstract}

We investigate the resilience of learning-based \textit{Intelligent Navigation Systems} (INS) to informational flow attacks, which exploit the vulnerabilities of IT infrastructure and manipulate traffic condition data. To this end, we propose the notion of \textit{Wardrop Non-Equilibrium Solution} (WANES), which captures the finite-time behavior of dynamic traffic flow adaptation under a learning process. The proposed non-equilibrium solution, characterized by target sets and measurement functions, evaluates the outcome of learning under a bounded number of rounds of interactions, and it pertains to and generalizes the concept of approximate equilibrium. Leveraging finite-time analysis methods, we discover that under the mirror descent (MD) online-learning framework, the traffic flow trajectory is capable of restoring to the Wardrop non-equilibrium solution after a bounded INS attack. The resulting performance loss is of order $\tilde{\mathcal{O}}(T^{\beta})$ ($-\frac{1}{2} \leq \beta < 0 )$), with a constant dependent on the size of the traffic network, indicating the resilience of the MD-based INS. We corroborate the results using an evacuation case study on a Sioux-Fall transportation network.
\end{abstract}


\section{Introduction} \label{intro}


The past decades have witnessed significant growth in the Internet-based traffic routing demand, along with the rapid development of modern \textit{Intelligent Navigation Systems} (INS). 
The INS infrastructures, which consists of Online Navigation Platforms (ONP) such as Google Maps and Waze, together with the widely adopted Internet-of-Things (IoT), including smart road sensors and toll gates, are designed to make real-time, efficient routing recommendations for their users. The best-effort routing of the individuals leads to macroscopic traffic conditions, which is 
encapsulated by the notion of \textit{Wardrop equilibrium} (WE) \cite{wardrop1952road} in \textit{congestion games}. 


As transportation networks become increasingly interconnected, the number of attack vectors against the entire transportation system is also on the rise. Consequently, the well-being of the traffic networks is vulnerable to emerging cyber-physical threats. For example, attacks on individual GPS devices and road sensors can lead to the unavailability of critical information and cause wide disruptions to the infrastructure.
As discussed in \cite{pan2022informational}, a strategic data poisoning attack on the ONP can lead to significant traffic congestion and service breakdown. 

In this work, we focus on a class of man-in-the-middle (MITM) attacks on ONP systems that aim to mislead the users to choose routes that are favored by the attackers.  
A quintessential case was demonstrated in 2014, Israel, where two students hacked the Waze GPS app and used bots to crowdsource false location information, which misled the users and caused congestion \cite{popularnavihack2014}. 
As reported in \cite{google2020}, the existing real-time traffic systems are intrinsically vulnerable to malicious attacks such as modified cookie replays and simulated delusional traffic flows. 

\begin{figure}
   \includegraphics[width=.45\textwidth]{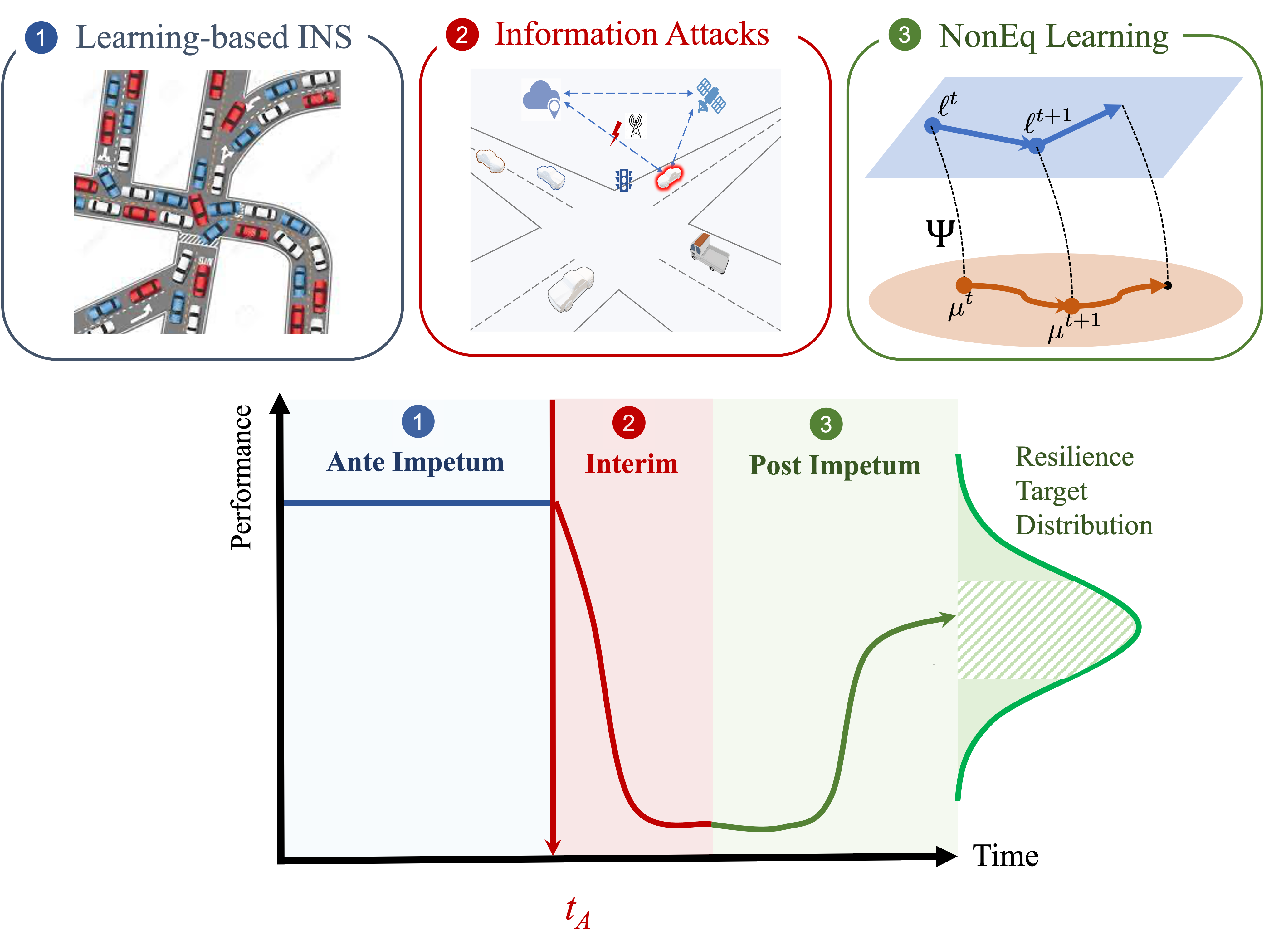}
   \caption{Intelligent Navigation Systems (INS) are vulnerable to informational attacks when data transmission is intercepted, yielding deteriorated traffic flows. Mirror descent, as a non-equilibrium learning scheme, is capable of restoring the path flow to the proposed Wardrop non-equilibrium solution.     }
   \label{examplefig}
\end{figure}




This class of attacks can be further categorized as {\it informational attacks} on INS. They  intercept the communication channel between individual users and the information infrastructure and exploit the vulnerabilities of data transmission to misguide users and achieve an adversarial traffic condition. 
While there have been efforts on preventing and detecting attacks, it is indispensable to create resilient mechanisms that can allow users to adapt and recover after the attack since perfect protection is either cost-prohibitive or impractical \cite{zhu2020cross,ishii2022security}. To achieve this self-healing property, a dynamic feedback-driven learning-based approach is essential \cite{huang2022reinforcement}, and a non-equilibrium solution concept in contrast to the classical Wardrop equilibrium is needed to capture the non-stationary nature of the {\it ante impetum} and {\it post impetum} behaviors as well as enable the time-critical performance assessment and design for resiliency.  



To this end, we investigate the notion of \textit{Non-Equilibrium Solution} (NonES) in the context of repeated congestion games.
It measures the probability of the traffic flows ``enveloping'' given \textit{target sets}, with the envelop volume defined by a \textit{measurement function}. The Non-Equilibrium learning does not necessarily yield an equilibrium solution but a trajectory that falls into the envelope with high probability.
Based on NonES, we define \textit{Wardrop Non-Equilibrium Solution} (WANES), which specifies mean Wardrop equilibrium (MWE) as its target set, and the weighted potential loss as the measurement.


In this work, we focus on a class of Mirror Descent Non-Equilibrium learning algorithms and elaborate on its role in  the resilience of traffic networks under adversarial environments.
We first establish a high probability bound on the distance between the output and MWE for generic \textit{Mirror Descent} (MD) algorithms without assumptions on the boundedness of the latency function. 
This high probability bound can be transformed into the resilience metrics, showing that after the attack, a WANES with weighted potential loss that is sublinear in time can be recovered through learning. 
Next, we develop a learning-based resiliency mechanism based on an MD algorithm as the and two classes of flow disturbance attacks.  We demonstrate the performance resilience under MD using an evacuation case study to illustrate the process of learning-based recovery. A schematic illustration of our non-equilibrium learning approach is provided in \Cref{examplefig}. 


{\bf Outline of this paper.} We briefly discuss the related works in Section \ref{rw}. 
In Section \ref{pf}, we introduce the repeated stochastic congestion game and MWE as the solution concept, based on which we introduce the notion of Non-Equilibrium learning and the formalism of resilience.
In Section \ref{ra}, we establish several finite-time results for the Non-Equilibrium learning dynamics and elaborate on the numerical experiments to illustrate the attack and resilience in Section \ref{sfdemo}.

\section{Related Work} \label{rw}
Our work bridges the gap between the online-learning and resilience in traffic assignment. Traffic assignment naturally fits in the online-learning framework, see \cite{blum2006routing,krichene2014convergence,krichene2015convergence}, where the convergence in Ces\'aro sense is shown as interests. Vu et al. in \cite{Vu2021} improved the rate of Ces\'aro convergence to $\mathcal{O}(1/T^2)$ and provided a last-iterate convergence guarantee.

The existent studies on the resilience of traffic assignment have targeted on event-based disruptions, (see, e.g., \cite{siri2020progressive}), the system impedance to misinformation disturbance has been rarely studied. To the best of our knowledge, the first informational equilibrium-poisoning concept was proposed by Pan et al. \cite{pan2022informational}, such a phenomenon occurs when the sensors, GPS devices in the INS are under attacks \cite{lou2016decentralization}. 

\section{Problem Formulation} \label{pf}

\subsection{Preliminary: Mean Wardrop Equilirbium}

We are given a traffic network represented as a directed, finite, and connected graph $\mathcal{G} = (\mathcal{V}, \mathcal{E})$ without self-loops. The vertices $\mathcal{V}$ represent road junctions, and the edges $\mathcal{E}$ represent road segments. The set of distinct origin-destination (OD) pairs is $\mathcal{W} \subseteq \mathcal{V} \times \mathcal{V}$, indexed by $w$, with cardinality $W$. Let $\mathcal{P}: = \bigcup_{w\in\mathcal{W}}\mathcal{P}_w$ be the set of all directed paths between origins and destinations, where $\mathcal{P}_w \subseteq \mathcal{P}(\mathcal{E})$ is the path set between the pair $w$. 

We assume that there is a set of infinitesimal players over $\mathcal{G}$, denoted by a measurable space $(\mathcal{X}, \mathcal{M}, m)$. 
The players are non-atomic, i.e., $m(x) = 0 \ \ \forall x \in \mathcal{X}$; they are split into distinct populations indexed by the OD pairs, i.e., $\mathcal{X} = \bigcup_{w\in\mathcal{W}} \mathcal{X}_w$ and $\mathcal{X}_w \bigcap \mathcal{X}_{w^{\prime}} = \emptyset,\ \forall w, w^{\prime} \in \mathcal{W}$. 
For each OD pair $w \in \mathcal{W}$, let $m_w = m(\mathcal{X}_w) $ represent the traffic demand. 
Let $\bar{M} = \sum_{w\in\mathcal{W}}m_w$ 
For each player $x \in \mathcal{X}_w$, we assume that their travel path $a \in \mathcal{P}_w$ is fixed right after the path selection.
The action profile of all the players $\mathcal{X}$ induces an edge flow vector $q \in \R^{|\mathcal{E}|}_{\geq 0}$, where $q_e := \int_{\mathcal{X}} \mathds{1}_{\{e \in a\}} m(dx), e \in \mathcal{E}$, and a path flow vector $\mu\in \Delta:=\{(\mu_p)_{p\in \cup_{w\in \mathcal{W}}\mathcal{P}_w} | \mu_p :={\int_{\mathcal{X}_w} \mathds{1}_{\{a = p\}} m(dx)}\}$.

We define the edge-path incident matrix of graph $\mathcal{G}$ as $\Lambda = [\Lambda^1 \vert , \ldots, \vert \Lambda^{|\mathcal{W}|}] \in \R^{|\mathcal{E}| \times |\mathcal{P}|}$ such that $\Lambda^w_{e, p} = \mathds{1}_{\{e\in p\}}, \forall e\in \mathcal{E}, w \in \mathcal{W}, p \in \mathcal{P}_w$.
Hence the compact form of edge-path flow relation is $q = \Lambda \mu$.

Let $(\Omega, \mathcal{F}, \mathbb{P})$ be the probability space for the formalism, $l_e: \R_{\geq 0} \times \Omega \mapsto \R_+$ be the cost/latency functions, measuring the travel delay of the edge $e \in \mathcal{E}$ determined by its edge flow $q_e$ and a state variable $\omega \in \Omega$ that is universal for the entire traffic network, e.g., $\omega$ can represent the weather condition, road incidents or anything that affects the congestion level. 
Let $l : \R^{|\mathcal{E}|}_{\geq 0} \times \Omega \mapsto \R^{|\mathcal{E}|}_+$ denote the vector-valued latency function. For an instance $\omega \in \Omega$, the latency of path $p$ is defined as $\ell_p : = \sum_{e \in p} l_e (q_e, \omega ) = \Lambda^{\top}_p l (\Lambda \mu, \omega )$, which can be seen as a function of $\mu$ and $\omega$, written as $\ell_p = \ell_p(\mu, \omega)$. 
We write the vector-valued path latency function as $\ell : \Delta \times \Omega \mapsto \R^{|\mathcal{P}|}_+$. Each instance $\omega$ determines a congestion game, captured by the tuple $\mathcal{G}_c^{\omega} = ( \mathcal{G}, \mathcal{W}, \mathcal{X} , \mathcal{P}, \mathcal{\ell}(\cdot, \omega) )$. Each path flow profile $\mu \in \Delta$ induces a probability measure associated with the positive random vector $\ell(\mu, \cdot): \Omega \mapsto \R^{|\mathcal{P}|}_+$.

\begin{assumption}\label{standingassumption} 
  For all $e \in \mathcal{E}$, the latency functions $l_e$ are $\omega$-measurable, for all $\omega \in \Omega$, $l_e$ are $L_0$-Lipschitz continuous and differentiable in $q_e$ with $\cfrac{ \partial l_e (q_e, \omega)}{\partial q_e} > 0 $ for all $q_e \geq 0$.
\end{assumption}
\begin{remark}
    The continuity assumption reflects the fact that adding a small amount of traffic does not drastically affect the travel latency; the monotonicity implies the increments of traffic does not decrease the latency.
\end{remark}

We adopt a stochastic alternative definition of Wardrop equilibrium. 
In doing so, 
we consider a ``meta'' version of the congestion game, $\mathcal{G}_c = (\mathcal{G}, \mathcal{W}, \mathcal{X}, \mathcal{P}, \mathbb{E}_{\omega}[\ell(\cdot, \omega)])$, with the utility functions replaced by the expected latency function. This ``meta'' congestion game gives rise to a solution concept corresponding to Definition \ref{wedef}. 

\begin{definition}[Mean Wardrop Equilibrium \cite{wardrop1952road}] 
\label{wedef}
 A path flow $\mu \in \Delta$ is said to be a \textit{Mean Wardrop Equilibrium} (MWE) if $\forall w \in \mathcal{W}$, $ \mu_p > 0$ indicates $ \E [\ell_p] \leq \E[\ell_{p^{\prime}}]$ for all $p^{\prime} \in \mathcal{P}_w$. 
 The set of all MWE is denoted by $\boldsymbol{\mu}^*$. 
 Equivalently, $\mu \in \boldsymbol{\mu}^*$ if and only if the following variational inequality is satisfied: $
    \mathbb{E}_{\omega} [\langle \mu - \mu^{\prime}, \ell(\mu, \omega) \rangle ] \leq 0  , \quad \forall \mu^{\prime} \in \Delta$,  where $\mathbb{E}_{\omega}[\cdot]$ is the expectation operator with respect to $\omega$.
\end{definition}

The MWE $\boldsymbol{\mu}^*$ is in general not a singleton, but a convex set given the strict monotonicity of latency in Assumption \ref{standingassumption}.
The seeking of MWE can be cast as a minimization problem of the expectation of the Stochastic Beckmann Potential (SBP) \cite{beckmann1956studies}, defined as $\phi(\mu, \omega) = \sum_{e \in \mathcal{E}}\int_{0}^{(\Lambda\mu)_e} l_e(z, \omega)dz$, where $(\Lambda \mu)_e$ is the $e$th element of $\Lambda \mu$. 
We refer to the expectation defined in \eqref{beckmanopt} as the Mean Beckmann Potential (MBP),
\begin{equation} \label{beckmanopt}
       \Phi (\mu) := \E \left[\sum_{e \in \mathcal{E}}\int_{0}^{(\Lambda\mu)_e} l_e(z, \omega)dz \right]. 
\end{equation}
It immediately follows that by Assumption \ref{standingassumption} that, for all $\mu \in \Delta$, 
\begin{equation*}
\begin{aligned}
        \nabla_{\mu} \Phi (\mu) & = \E [ \Lambda^{\top} l (\Lambda \mu, \omega)] = \E[\ell (\mu, \omega )],  \\
        \nabla^2_{\mu} \Phi (\mu) & = \E [ \Lambda^{\top} (\nabla l ( \Lambda \mu, \omega)) \Lambda ] \succeq  0.
\end{aligned}
\end{equation*}
Therefore, $\Phi$ is convex in $\mu$. The characterization of the MWE coincides with the first-order optimality condition. Finally, we denote by $\Phi^*$ the unique optimal BMP: $\Phi^* := \min_{\mu \in \Delta} \Phi(\mu)$.


\subsection{ Mirror Descent and Wardrop Non-Equilibrium}

In the online-learning setting, the players make decisions repeatedly.
Let the time index be $t \in \mathbb{N}_+$, for each OD pair $w$,  each player $x \in \mathcal{X}_w$ receives a mixed strategy $\pi^t (\cdot, x): \mathcal{X}_w \mapsto \Delta (\mathcal{P}_w)$ which is $\mathcal{M}$-measurable,
and plays a randomized routing path $A^t(x) \sim \pi^t ( \cdot ,x)$. 

Under identical and independent path choice randomization within the populations of each OD pair, individual-level and population-level online learning are equivalent due to the non-atomic nature, \cite{krichene2014convergence}. 
We hereby let the history $(\mathcal{H}_t)_{t \geq 0}$ be a sequence of realizations of $\omega^t$, $\ell^t$, and $\mu^t$ up to time $t$, an online-learning algorithm $\mathcal{A}$ maps from the space of $\mathcal{H}_t$ to $\Delta$, iteratively generating the traffic flow $\mu^{t+1}$.

The individual regret with respect to a path choice $p \in \mathcal{P}_w$ for $x \in \mathcal{X}_w$, $w \in \mathcal{W}$ is $\mathcal{R}_T (x) = \mathbb{E}[\sum_{t=1}^T \ell^t_{A^t(x)} - \ell_p^t]$, where $\mathbb{E}[\cdot]$ is taken with respect to $\mathcal{A}$ and $\omega$.
The population regret with respect to a path flow is defined as $ \mathcal{R}_T (\mu) = \mathbb{E}[\sum_{t = 1}^T \langle \mu^t - \mu, \ell(\mu^t, \omega^t) \rangle]$. 
Let $\mu^*$ be one of $\boldsymbol{\mu}^*$, by convexity of $\Phi$, a sub-linear regret bound, i.e., $\mathcal{R}_T(\mu^*) =  o(T) $ directly implies $\Phi(\bar{\mu}^T) \to \Phi(\mu^*)$ as $T \to \infty$, where $\bar{\mu}^T := \frac{1}{T}\sum_{t=1}^T \mu^t$ is the empirical flow. 

To achieve such sub-linear regret bound [$\mathcal{R}_T(\mu^*) =  o(T)$], a class of widely used online learning algorithms can be obtained through \textit{mirror descent} (MD), as shown in Algorithm \ref{omdalg}, with a specified instance of Bregman divergence $D_{\Psi}(\cdot, \cdot): \Delta \times \Delta \mapsto \R$.  Induced by a mirror map $\Psi: \Delta \mapsto \bar{\R} $, the divergence $D_{\Psi}(\mu_1 , \mu_2) := \Psi(\mu_1) - \Psi(\mu_2) - \langle \nabla \Psi(\mu_2), \mu_1 - \mu_2 \rangle$ measures the dissimilarity between two iterates [see \eqref{mddyna}], regularizing the learning process.   

The mirror map $\Psi$ is assumed to be Fr\'echet differentiable and strongly convex, i.e., there exists a constant $\sigma_{\Psi} > 0$ such that $D_{\Psi} (\mu_1, \mu_2)  \geq \frac{\sigma_{\Psi}}{2} \| \mu_1 - \mu_2 \|^2$. Note that when the mirror map is given by $\ell_2$-norm, mirror descent in \eqref{mddyna} reduces to projected gradient descent \cite{nesterov04book}: $ \mu^{t+1}\leftarrow \argmin_{\mu\in \Delta}\|\mu^t+\eta_t\ell_t-\mu\|^2$. Hence, MD is a generalization of gradient methods and allows more freedom when designing learning algorithms. For example, when $\Psi(\mu)=\frac{1}{2}\|\mu\|_p^2, 1<p<2$, mirror descent works favorably for sparse problems \cite{lei18md-hp}.

\begin{algorithm}
\SetKwInOut{Input}{Input}
\Input{initialize $\mu^1 \in \Delta$, learning rate $(\eta_t)_{t \in \mathbb{N}_+}$.}
\For{$t \in \mathbb{N}_+$}{
     \For{$w \in \mathcal{W}$, $x \in \mathcal{X}_w$, } 
     { 
     INS assigns mixed strategy $\pi^t(\cdot, x) \leftarrow \frac{1}{m_w} (\mu^t_p)_{p \in \mathcal{P}_w}$ to player $x$\;
     player $x$ samples path $A(x) \sim \pi^t(\cdot, x)$;
     }
     nature samples $\omega^t \sim \mathbb{P}(\cdot)$\;
     INS reveals latency vector $\ell^t = \ell (\mu^t, \omega^t)$ to $\mathcal{X}$\;
     INS updates:
     \begin{equation}\label{mddyna}
         \mu^{t+1}  \leftarrow \argmin_{\mu \in \Delta}  \eta_t \langle  \mu , \ell^t \rangle + D_{\Psi}(\mu, \mu^t)
     \end{equation}
}
\caption{\texttt{Mirror Descent} for INS}
\label{omdalg}
\end{algorithm}
It is shown in \cite{krichene2015convergence} that the populational regret under MD  can achieve $\mathcal{O}(\sqrt{T})$ in the static regime (when $\Omega$ is a singleton), which coincides with the $\tilde{\mathcal{O}}(\sqrt{T})$ results shown for stochastic environment in \cite{Vu2021}. These sublinear bounds suggest that the empirical flow under MD arrives at the MWE asymptotically.


However, the asymptotic convergence of empirical flow to the MWE does not capture the transient behavior of the learning process. The  regret bounds above does not answer the following question regarding the resiliency of MD: \textit{how many iterates does MD need to recover from an informational attack and return an approximate MWE?}
The insufficiency of asymptotic equilibrium characterization motivates us to dive into the finite-time analysis of the learning process. Instead of studying the limiting behavior of learning iterates, we shift the focus
to finite sequences of iterates and associated probabilistic characterizations, based on which we propose a new solution concept for learning algorithms: \textit{Non-Equilibrium Solution} (NonES). For congestion games, NonES is captured by a measurement function and a target set of flow profiles.  A finite sequence of iterates (called a
trajectory) produced by the learning algorithm $\mathcal{A}$ is treated as a random variable whose probability measure is determined by
the learning dynamics. Then, the measurement function maps this random variable to the space where the target set is defined. Whether the trajectory (transformed by the measurement function) falls within
the target set constitutes a random event. The probabilistic characterization of this random event is the basis of the non-equilibrium definition introduced in the following. 
\begin{definition}[Non-Equilibrium Solutions]
\label{noneqgeneric}
For a congestion game $\mathcal{G}_c$, denoted by $\mathcal{B}$ the Borel $\sigma$-algebra over the set of path flow profiles $\Delta$. Let $\Delta^t$ and $\mathcal{B}^t$ be the product space and the product measure, respectively. Denote by $\mathbb{P}_t$ the probability measure over the space $(\Delta^t, \mathcal{B}^t)$. Given a measurement function $F:\Delta^t\rightarrow \Delta$, a target set $\mathcal{C}\subset \Delta$, and a positive number $\delta>0$, $\mathbb{P}_t$ is an $(F,\mathcal{C},\delta)$-Non-Equilibrium solution (NonES) if 
\begin{align}
    \mathbb{P}_{t}\{(\mu^k)_{k=1}^t\in \Delta^t| F[(\mu^k)_{k=1}^t]\in \mathcal{C}\}\geq 1-\delta.\label{eq:nones}
\end{align}
\end{definition}
\begin{remark}[Non-Equilibrium Learning]
    In the context of online learning, the probability measure $\mathbb{P}_t$ is determined by the learning algorithm $\mathcal{A}$ and the environment stochasticity $\omega_t$ in Algorithm \ref{omdalg}. We refer to $\mathcal{A}$ as Non-Equilibrium learning if the induced probability measure $\mathbb{P}_t$ is a NonES, as formally defined in \Cref{wanes}. When no confusion arises, we also say a trajectory $(\mu^k)_{k=1}^t$ under $\mathcal{A}$ is a NonES if its associated probability measure is a NonES. 
\end{remark}

The proposed non-equilibrium solution generalizes existing equilibrium-seeking characterizations (e.g., last iterate convergence and Ces\'{a}ro convergence in \Cref{exam}), which concerns transient properties of the underlying flow sequence. A probability measure (or equivalently a distribution) over the sequences of flows is a NonES if the sequences fall within the target set with high probability ($\delta>0$) or almost surely ($\delta=0$).  
Our resilience study of online learning algorithms is built upon this Non-Equilibrium notion, where we demonstrate that MD can quickly recover from unexpected perturbation, and the resulting path flow falls within a neighborhood of the optimal one.  The following example shows that the widely used  Ces\'{a}ro convergence \cite{krichene2014convergence,krichene2015convergence} is a special case of the proposed Non-Equilibrium.

\begin{example}
\label{exam}
A sequence of path flows $\{\mu^k\}_{k=1}^\infty$ is said to Ces\'{a}ro converge to  MWE  with respect to weights $\{\eta_k\}_{k=1}^\infty$ almost surely if $\lim_{t\rightarrow\infty}\Phi(\bar{\mu}^t)=\Phi^*, \bar{\mu}^t=\sum_{k=1}^t \eta_k\mu^k/\sum_{k=1}^t \eta_k$, with probability $1$. When $\eta_k=1$, for all $k\geq 1$, the Ces\'aro average reduces to the empirical flow: $\bar{\mu}^t=\frac{1}{t}\sum_{k=1}^t \mu^k$. Note that in the online learning context, the weights $\{\eta_k\}_{k=1}^\infty$ correspond to the vanishing learning rates of some algorithm $\mathcal{A}$ (e.g., learning rate in Algorithm~\ref{omdalg}).

The following rephrases the convergence characterization above using non-equilibrium language. Given an MWE $\mu^*$, for any $\epsilon>0$, let the target set $\mathcal{C}_\epsilon$ be the $\epsilon$-approximate MBE, i.e., $\mathcal{C}_\epsilon:=\{\mu\in \delta| \Phi(\mu)-\Phi^*<\epsilon\}$.  Define the measurement function as the weighted Ces\'{a}ro average, $F[(\mu^k_{k=1})^t]=\sum_{k=1}^t \eta_k\mu^k/\sum_{k=1}^t \eta_k$. A sequence of flows $\{\mu^k\}_{k=1}^\infty$ converges to MWE if for any $\epsilon>0$, there exists a $T$ such that any finite subsequences $\{\mu^k\}_{k=1}^t, t>T$ is an $(F, \mathcal{C}_\epsilon,0)$-Non-Equilibrium.
\end{example}

Introducing the target set and measurement function provides additional degrees of freedom when analyzing the transient behavior of a sequence of flows produced by learning processes. For example, the measurement function can be defined as the last iterate MBP returned by the learning algorithm. In this case, the proposed NonES generalizes the way to characterize the outcome of the last iterate  \cite{Vu2021}. 

\begin{definition}[Wardrop Non-Equilibrium Learning]
\label{wanes}
For the congestion game $\mathcal{G}_c$, let the measurement function be the Ces\'{a}ro average in \cref{exam}. For any $\epsilon>0$, define the target set as $\mathcal{C}_\epsilon:=\{\mu\in \Delta| \Phi(\mu)-\Phi^*<\epsilon\}$. A probability measure $\mathbb{P}_t$ over $(\Delta^t, \mathcal{B}^t)$ is an $(\epsilon,\delta)$-Wardrop Non-Equilibrium solution (WANES) if  $\mathbb{P}_t\{(\mu^k)_{k=1}^t\in \Delta^t| \bar{\mu}^t\in \mathcal{C}_{\epsilon}\}\geq 1-\delta.$
Furthermore, any learning algorithm $\mathcal{A}$ producing such $\mathbb{P}_t$ is said to be an $(\epsilon,\delta)$-Wardrop Non-Equilibrium learning. 
\end{definition}

\subsection{Resilience to Informational Attacks}
\label{sec:resilience}
While under normal operation, the INS traffic flow is close to the equilibrium flow set $\boldsymbol{\mu}^*$, a one-shot perturbation in the flow can cause successive disruptions, as discussed in Section \ref{intro}. 
To see this, let the actual flow of the transportation network be $\mu^{t_0} \in \boldsymbol{\mu}^*$ at time $t_0 \in \mathbb{N}_+$, an MITM attacker is able to modify this piece of information into $ \mu^{\dagger} \in \Delta$ to mislead the INS. This error in turn propagates to the latency vector revealed by the INS so that at time $t$ the loss vector $\ell^t$ is replaced by $\tilde{\ell}^t : = \ell(\mu^{\dagger}, \omega^t)$.
Hence the mirror step \eqref{mddyna} at $t_0$ is poisoned as follows:
\begin{equation}\label{poisonedmdstep}
     \mu^{t_0+1} \leftarrow \argmin_{\mu \in \Delta} \eta_t \langle  \mu , \tilde{\ell}^t \rangle + D_{\Psi}(\mu, \mu^{\dagger}).
\end{equation}

The INS assigns  the individual mixed strategies corresponding to poisoned $\mu^{t_0+1}$, hence propagating the flow disturbance attack.
Let attack $a^{\dagger} := D_{\Psi}(\mu^{t_0 }, \mu^{\dagger})$ be the Bregman divergence from $\mu^{\dagger}$ to $\mu^{t_0}$, which stands for the attack magnitude in terms of the information geometry, as the flows can be scaled as probability distributions. 

However, the intrinsic adaptability of MD enables the INS to pull the poisoned flow back to the right track, by iterative Non-Equilibrium learning in the environment. We hereby give a resilience characterization for such adaptability in Definition \ref{resiliencedef} based the non-equilibrium notion.

\begin{definition}[Resilience]\label{resiliencedef}
Given an attack $a^{\dagger} \in \R_+$, let $r_{a^{\dagger}}(\cdot, \cdot): (0,1) \times \mathbb{N}_+ \mapsto \R_+$ be a recovery threshold function parameterized by $a^{\dagger}$, $T \in \mathbb{N}_+$ be a recovery time length, and $\mathcal{A}$ be an online-learning algorithm.
For $\delta \in (0,1)$, the INS is said to be $( r_{a^{\dagger}}, T, \delta)$-resilient under $\mathcal{A}$ if the $T$-step trajectory $(\mu^1, \ldots, \mu^T)$ under $\mathcal{A}$ after an attack $a^{\dagger}$ is a $(r_{a^{\dagger}}, \delta)$-WANES, i.e., $ \mathbb{P} \left\{\Phi(\bar{\mu}^T)-\Phi^*< r_{a^\dagger}\right\} \geq 1 - \delta$.

\end{definition}

The resilience of the INS is quantified by the ability to recover from a given attack $a^{\dagger}$. It is natural that the ability to recover is dependent on the level of $a^{\dagger}$ and the recovering time $T$, with $\delta$ picked as a tolerance parameter for recovery-failing tail probability.

\section{Resilience Analysis} \label{ra}

\subsection{Resilience with General $\Psi$}

 We introduce the following simplified notations. Let the optimal SBP be $\phi^*(\omega) := \sup_{\mu \in \boldsymbol{\mu}^*}\phi(\mu, \omega)$, which is assumed to be finite almost surely; the worst case potential is then $\phi^* = \sup_{\omega\in \Omega} \phi^*(\omega)$; the realized SBP at time $t$ is $\phi^t = \phi(\mu^t, \omega^t)$; the MBP at time $t$ be $\Phi^t := \Phi(\mu^t )$. Let $d(\mu, \boldsymbol{\mu}^*) = \inf_{\mu^* \in \boldsymbol{\mu}^*} \| 
 \mu - \mu^*\|^2$ be the Euclidean distance from $\mu$ to the set of MWE.  
We impose a standard technical assumption that significantly simplifies the analysis. 

\begin{assumption} \label{lineargrowthassume}
    For all $\mu \in \Delta$ $\omega \in \Omega$, the latency function satisfies that, there exist two constants $A$ and $B$ such that$
         \|\ell(\mu, \omega) \|^2 \leq A \phi(\mu, \omega) + B$.
\end{assumption}
Assumption \ref{lineargrowthassume} indicates the linear growth of $\| \ell \|^2$ with respect to $\phi$, which can be analytically verified for some particular choices of Bureau of Public Roads (BPR) function, e.g., the additively perturbed BPR functions of form $l^{\bf bpr}_e((\Lambda \mu)_e, \omega ) = t_e (1 + \alpha_1 (1 + \frac{(\Lambda \mu)_e}{C_e})^{\alpha_2}) + \omega_e$, where $\omega \in \Omega \subseteq \R^{|\mathcal{E}|}$ is the edge-wise perturbation vector, $t_e$ is free travel time, $C_e$ is the edge capacity, with $\alpha_1$ and $\alpha_2$ being two parameters.  We begin our analysis with Lemma \ref{distanceofbregman}, which bounds the one-step change of $D_{\Psi}(\mu, \mu^t)$ by MD.

\begin{lemma}\label{distanceofbregman}
Let $\{\mu^t\}_{t\in \mathbb{N}}$ be a sequence generated by \eqref{mddyna}, then the following holds for any $\mu \in \Delta$, 

\begin{equation} \label{onestepinequal}
\begin{aligned}
 & \quad D_{\Psi}\left(\mu, \mu^{t+1}\right) -D_{\Psi}\left(\mu, \mu^t\right)  \\ & \leq \eta_{t}\left\langle \mu-\mu^{t}, \ell^t \right\rangle  + 2 \frac{\eta_{t}^{2} }{\sigma_{\Psi}} (A \phi^t+B)
\end{aligned}
\end{equation}
\end{lemma}

$(A\phi^t + B)$ term can be replaced with a coarser bound, but this refined one-step inequality \eqref{onestepinequal}, as an outcome of Assumption \ref{lineargrowthassume}, gives a profound interpretation as it connects the divergence change to $\phi^t$. This divergence difference characterizes the system-level ``rationality'': as $\phi^t$ gets lower, the ``rationality'' level gets higher, and less effort needs to be paid to change the flow.

Based on Lemma \ref{distanceofbregman}, Lemma \ref{distanceofmus} bounds the Euclidean distance from post-attack $\mu^t$ to $\boldsymbol{\mu}^*$. To  simplify the analysis, set $t_0 = 1$ by default and assume that at $t_0$, the INS already reaches the WE set, i.e.,  $\mu^{t_0} \in \boldsymbol{\mu}^*$. The attacker launches $a^{\dagger} = D_{\Psi}(\mu^{t_0}, \mu^{\dagger}) \geq \inf_{\mu \in\boldsymbol{\mu}^*} D_{\Psi} (\mu , \mu^{\dagger})$, after which the MD dynamic initializes $\mu^1 \leftarrow \mu^{\dagger}$.

\begin{lemma}\label{distanceofmus}
    Let $\{\mu^t\}_{t \in \mathbb{N}_+}$ be a sequence generated by \eqref{mddyna} after the attack  $a^{\dagger}$, let $C_1 =  \phi^* + \frac{B}{A}$, with $\eta_t \leq \frac{\sigma_{\Psi}}{2 A}$ and being non-increasing, we have for all $t \in \mathbb{N}_+$, $\mu^* \in \boldsymbol{\mu}^*$,
    \begin{align} \label{dismus}
        d (\mu^{t+1}, \boldsymbol{\mu}^*)   \leq     \|\mu^{t+1} - \mu^* \|^2 \leq 
 2\sigma_{\Psi}^{-1} (C_1  \sum_{k=1}^{t}\eta_k +  a^{\dagger}) 
 \end{align}
and the following upper bounds: $\sum_{k=1}^{t} \eta_{k}^{2}\phi^k    \leq 2 (C_{1} \sum_{k=1}^{t} \eta_{k}^{2} + \eta_1 a^{\dagger})$ and $\sum_{k=1}^{t} \phi^k   \leq 2( C_{1} t+  ( C_{1}\sum_{k=1}^{t}  \eta_{k} + a^{\dagger}) \eta_{t}^{-1} + \eta_1^{-1} a^{\dagger})$.

\end{lemma}

Lemma \ref{distanceofmus} gives a distance bound larger than $2 \sigma_{\Psi}^{-1} a^{\dagger}$, increasing with $t$, yet allows us to control the distance by adjusting the order of the summation $\sum_{k=1}^t \eta_k^2$, which is convergent as $t \to \infty$ under careful tuning, e.g., when $\eta_t = \eta_1 t^{\beta - \frac{1}{2}}$ with $\beta \in (-\frac{1}{2}, 0)$.
Later we show that, $d(\mu^{t+1}, \boldsymbol{\mu}^*)$ can be controlled by the $\mathcal{O} (\sum_{k=1}^t \eta_k^2 d (\mu^k, \boldsymbol{\mu}^*))$ with high probability, which allows us to bound the maximum of $d(\mu^{t+1}, \boldsymbol{\mu}^*)$, 
as stated in Theorem \ref{distancemuwhp}.

\begin{theorem}\label{distancemuwhp}
    Let $\{\mu^t\}_{t\in \mathbb{N}_+ }$ be the sequence generated by \eqref{mddyna} after attack $a^{\dagger}$, assuming that $\eta_t \leq \frac{\sigma_{\Psi}}{2 A}$ and is non-increasing. Let the two quantities be $c_1 := \max_{k\in \mathbb{N}_+} \eta_k \sum_{j=1}^{k-1}\eta_j < \infty$, $c_2 := \eta_1 ( \phi^* + A \Phi^* + B) + \sigma^{-1} (2A^2 + 1)( C_1c_1  + \eta_1 a^{\dagger})$, we have for $t \in \mathbb{N}_+$, for $\delta \in (0,1 )$, w.p. $1 - \delta$, 
    \begin{equation}
     \max_{1 \leq t \leq T}d( \mu^t, \boldsymbol{\mu}^*) \leq C_2 \log \left(\frac{T}{\delta} \right),
    \end{equation}
    where $C_2=\frac{4 c_2  }{\sigma_{\Psi}\rho} + ( \frac{ 4}{\sigma_{\Psi}}  + \frac{8\eta_1 A}{\sigma_{\Psi}^{2}})a^{\dagger}   +  \frac{8AC_1 +4 B }{\sigma_{\Psi}^2}  \sum_{k=1}^t \eta_k^2++ \frac{\sum_{k=1}^{t_1} 2 C_1 \sum_{j=1}^{k-1}\eta_j + a^{\dagger}}{\sigma_{\Psi}(\frac{\eta_1}{C_1} a^{\dagger} + c_2) } $. 
\end{theorem}

Intuitively, the traffic flow output by MD should fall into a logarithmic ball centralized around the MWE flows, with the diameter dependent on the initial flow disturbance attack.
Based on Theorem \ref{distancemuwhp}, we can establish the high-probability resilience results as stated in Proposition \ref{resilienceprop}.

\begin{prop}[Resilience of MD]\label{resilienceprop}
Under attack $a^{\dagger}$,  let $\delta \in (0,1)$ and $r_{a^{\dagger}}(\cdot, \cdot)$ be defined as $r_{a^{\dagger}} (\delta, T)=  (\sum_{k=1}^T \eta_k)^{-1}C_3 (a^{\dagger})\log^{\frac{3}{2}} (\frac{2T}{\delta})$,
    where
    \begin{align*}
        C_3 (a^{\dagger})& = \bigg( \left(1 + \frac{2A \eta_1}{\sigma_{\Psi}}a^{\dagger}\right) + ((4A^2+ 1)C_2 + 4AC_1) \\& (2\sum_{t=1}^{\infty}\eta_t^2)^{\frac{1}{2}}
     + \frac{2(AC_1 + B)}{\sigma_{\Psi}} \left(\sum_{k=1}^{\infty} \eta_k^2\right) \bigg)
    \end{align*}
is a constant independent of $T$ but dependent on $a^{\dagger}$, with $C_2$ defined in Theorem \ref{distancemuwhp}.
The INS is $(r_{a^{\dagger}}, T, \delta)$-resilient under MD algorithm \ref{omdalg}. Furthermore, with $\eta_t = \eta_1 t^{-\beta-1}$ and $\beta \in (-\frac{1}{2}, 0)$, then $r_{a^{\dagger}}(\delta, T) = \mathcal{O}(T^{\beta} \log^{\frac{3}{2}} \frac{T}{\delta})$.
\end{prop}

In Proposition \ref{resilienceprop} we give a sub-linear order for the threshold function $r_{a^{\dagger}}$, without imposing boundedness assumption on the latency vector.
The $\tilde{\mathcal{O}}(T^{\beta})$ order implies a.s. convergence of the MBP to the optimum, which indicates the asymptotic collapse of performance loss. In the long run, the INS is expected to recover fully from such attacks. 

\subsection{Resilience Discussion with Bounded Attack}
In this section, we let $\Psi$ be the unnormalized negentropy, i.e., for $\mu \in \Delta$, $\Psi(\mu) = \sum_{p \in \mathcal{P}} \mu_p \log \mu_p - \mu_p$. In this case, the MD step gives, under the initial information disturbance $a^{\dagger}$, $\mu^{t+1}_p \propto \mu_p^t \exp( - \eta_t \tilde{\ell}_p^t)$ $ p\in \mathcal{P}_w$ $ w \in \mathcal{W}$, that is,  for all $w \in \mathcal{W}$,
\begin{equation} \label{mirrorstepwithentropy}
 (\mu^{t+1}_p)_{p \in \mathcal{P}_w} =  \left(m_w \frac{e^{- \sum_{s \leq t} \eta_s \tilde{\ell}^s_p}}{\sum_{p \in \mathcal{P}_w } e^{- \sum_{s \leq t} \eta_s  \tilde{\ell}^s_p} }\right)_{p \in \mathcal{P}_w},
\end{equation}
where $\tilde{\ell}^s$ is the feedback latency vector of the post-attack flows. To illustrate the dependence of resilience on the attack capacity, we consider two types of attacks:  \texttt{Unif} attacks and  \texttt{Supp} attacks. The \texttt{Unif} attacks are when $a^{\dagger}$ is such that $\mu^{\dagger}_p = \frac{1}{|\mathcal{P}_w|} m_w$, for all $p \in \mathcal{P}_w$, $w \in \mathcal{W}$, in which case the flow information is uniformly redistributed. 
The \texttt{Supp} attacks represent a more generic class of attacks, where the attacker poisons the flow information such that $ \mathrm{supp}(\mu^{t_0}) \subseteq \mathrm{supp}(\mu^{\dagger}) $. Both types of attacks satisfy the boundedness, i.e., $a^{\dagger} < \infty$.

\begin{prop} \label{examprop}
Let $ P = \max_{w \in \mathcal{W}} |\mathcal{P}_w|$, $\gamma_w = \min_{p \in\mathcal{P}_w, \mu_p > 0} \mu^{\dagger}_p $, and $\gamma = \min_{w \in \mathcal{W}}\gamma_w$, under MD algorithm \ref{omdalg}
\begin{itemize}
    \item[a)]  Under \texttt{Unif} attack, the INS is $(r_{\texttt{Unif}}, T, \delta)$-resilient, with $r_{\texttt{Unif}}(\delta, T) = \tilde{\mathcal{O}}( M W T^{\beta} \log P )$ for $\beta \in (-\frac{1}{2}, 0 )$.
    \item[b)] Under \texttt{Supp} attack, the INS is $(r_{\texttt{Supp}}, T, \delta)$-resilient, with $r_{\texttt{Supp}} (\delta, T) = \tilde{\mathcal{O}} (M \bar{M}^2 \gamma^{-1} T^{\beta})$  for $\beta \in (-\frac{1}{2}, 0 )$.
\end{itemize}
 \end{prop}

By Proposition \ref{examprop}, under \texttt{Unif} attack,  the resilience threshold is up to a logarithmic order of the maximum path size $P$, and linear in terms of the number of OD pairs. The  capability of the MD learning to adapt and recover the system is then linearly dependent on the network size $|\mathcal{E}| = \mathcal{O}(\log(|\mathcal{P}|))$ and the population complexity $M$ and $W$.
However, when the attack $a^{\dagger}$ becomes more random, the INS becomes less resilient as the $T^{\beta}$ order may be partially offset by other factors such as $\gamma$, slowing down the recovery.

\section{Case Study} \label{sfdemo}
This section studies an experimental setup of an evacuation process in Sioux Falls, SD, building on the South Dakota Transportation Network \cite{siouxfalls}.
At each time unit, a fixed number of individuals are transported from a set of emergency locations to shelter places. 
We adopt the BPR function discussed in Section \ref{ra} to generate the latency feedback. The transportation network data, including OD demand, free travel time, and road capacities are obtained from \cite{github}. 

\begin{figure}[htbp]
    \centering
    \includegraphics[width= .45\textwidth]{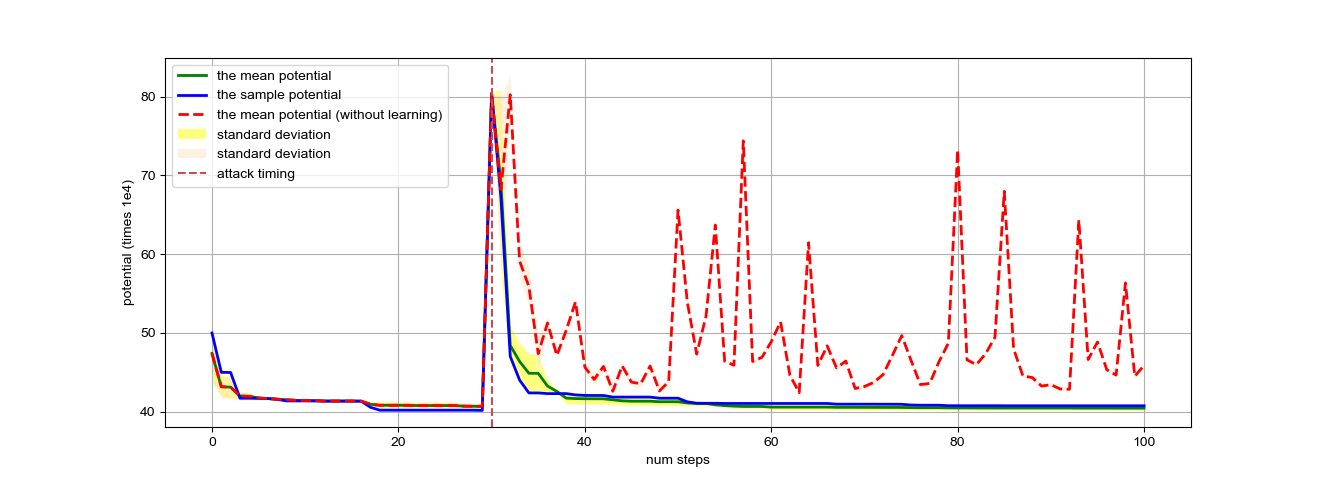}
     \caption{The spike at $t_0 = 30$ indicates the \texttt{Unif} flow disturbance. The red dotted curve represents the greedy assignment process; the blue and green curve represents the non-equilibrium learning process.}
     \label{resultfig}
\end{figure}

We assume that the evacuation process is conducted using a learning-based mechanism, i.e., the MD algorithm.  We simulate the learning process for $100$ time units, at $t_0 = 30$, the  we run the simulation for $10$ times and plot the mean and a sample of the MBP trajectory in Fig. \ref{resultfig}.

 As shown in the figure, at time $t_0 = 30$, an attacker launches a \texttt{Unif} attack on the INS, causing the potential to be much higher. After the attack, we compare the learning-based resilience and the recovery without learning by setting the benchmark as a greedy assignment process, which iteratively allocates a half portion of traffic demand to the path with minimum latency. In comparison with the greedy assignment, which produces potential oscillation after the attack, the INS can rapidly recover the system from high MBP through MD learning within $15$ time steps.
 
\begin{figure}[htbp] 
\centering
 \includegraphics[width =.45\textwidth]{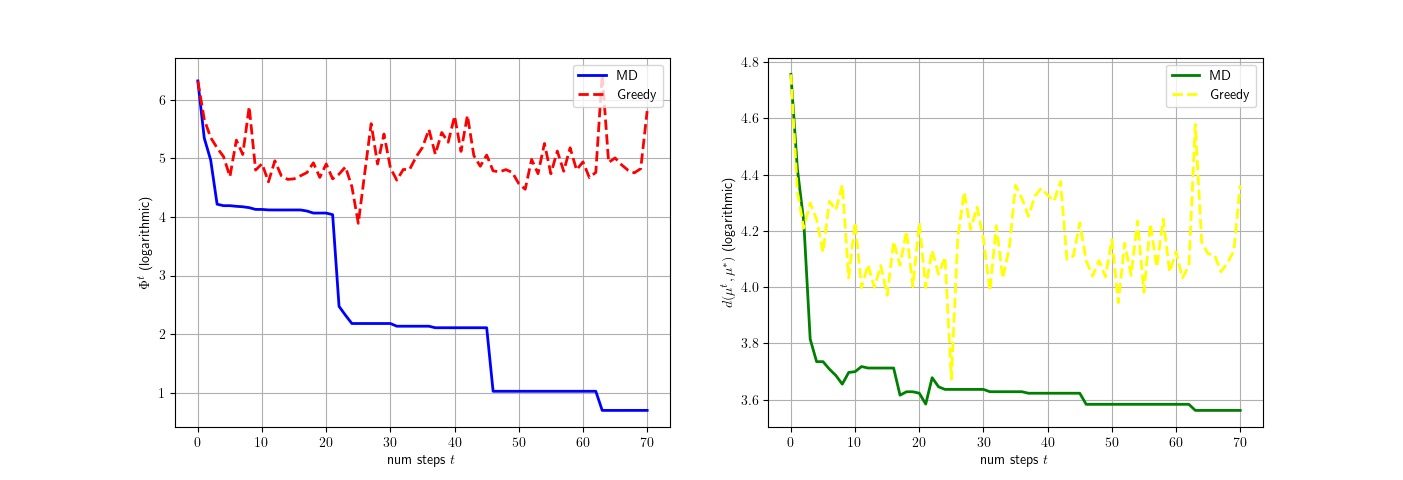}
 \caption{The two figures compare the learning-based (MD) resilience and greedy-assignment (Greedy) resilience, by showing the post-attack change of $\Phi^t - \Phi^*$ (left), and $d(\mu^t, \boldsymbol{\mu}^*)$ (right) over time respectively.}
 \label{logscale}
\end{figure}
By plotting the post-attack curves of MBP difference $\Phi^t - \Phi^*$ and $d(\mu^t, \boldsymbol{\mu}^*)$ with logarithmic order, as shown in Fig. \ref{logscale},  one can observe that the learning-based trajectory achieves faster recovery and better stability,  corroborating that the MD-based INS provides stronger resiliency and higher efficiency.

\section{Conclusions and Future Work}

In this paper, we have investigated the resilience of traffic networks under misinformation attacks on the Intelligent Navigation Systems (INS). The proposed non-Equilibrium learning has enabled a feedback-enabled resiliency mechanism and provided post-attack  resiliency assessment and design methodologies. 
Through finite-time analysis of the learning dynamics, we have demonstrated the ability of INS to recover from multiple informational attacks. Future research would focus on creating scalable and distributed resilience mechanisms that can scale up with respect to the time and network size of the transportation networks. We would investigate the dynamic attack model to develop defensive strategies against strategically evasive cyber-physical threats.

\bibliographystyle{IEEEtran}
\bibliography{ref}

\appendix 

\subsection{Two Technical Lemmas}

\begin{lemma}[Concentration Bounds \cite{zhang2005data}] \label{concentrationtech}
Let $\omega_1,\ldots,\omega_k$ be a sequence of random variables (not necessarily i.i.d.), let functionals $\xi_k(\omega_1,\ldots,\omega_k)$, $k = 1, \ldots, T$ be such that the conditional variance sum can be bounded $\sum_{k=1}^{T} \mathbb{E}_{\omega_{k}}\left[\left(\xi_{k}-\mathbb{E}_{\omega_{k}}\left[\xi_{k}\right]\right)^{2}\right] \leq \sigma^2$: 1) if
 if $
|\xi_{k}-\mathbb{E}_{\omega_{k}}\left[\xi_{k}\right] | \leq b_k$ for each $k$, for $\delta \in (0,1)$,
$
\mathbb{P} \left\{ \sum_{k=1}^{T} \xi_{k}-\mathbb{E}_{\omega_{k}}\left[\xi_{k}\right] \leq\left(2 \sum_{k=1}^{T} b_{k}^{2} \log \frac{1}{\delta}\right)^{\frac{1}{2}} \right\} \geq 1- \delta
$;
2) if $
\xi_{k}-\mathbb{E}_{\omega_{k}}\left[\xi_{k}\right] \leq b$, for each $k$
for $\rho \in (0,1)$, 
$\mathbb{P}\left\{
\sum_{k=1}^{T} \xi_{k}- \mathbb{E}_{\omega_{k}}\left[\xi_{k}\right] \leq \frac{\rho \sigma^{2}}{b}+\frac{b \log \frac{1}{\delta}}{\rho}
\right\} \geq 1- \delta$.

\end{lemma}

\begin{lemma} \label{boundphik}
    Let there be an arbitrary real-valued convex differentiable function in the domain $\Delta$, taking the MBP $\Phi$ for example, suppose it satisfies Assumption \ref{lineargrowthassume}, then, for all $\mu \in \Delta$, $\mu^*\in \boldsymbol{\mu}^*$, 
    $
       \|\nabla \Phi( \mu )\|^{2} = \|\mathbb{E}_{\omega} [\ell(\mu, \omega)]\|^2  \leq 2 (A^{2}\| \mu - \mu^*\|^{2}+A \Phi^*+ B)$.
\end{lemma}

\proof[Proof of Lemma \ref{boundphik}]{
By convexity of $\Phi$ and Assumption \ref{lineargrowthassume}, for any $\mu^* \in \boldsymbol{\mu}^*$,
\begin{equation*}
    \begin{aligned}
        \| \nabla \Phi(\mu)\|^2 & \leq A(\Phi(\mu)- \Phi^*) +A \Phi^* + B \\ 
        & \leq A \langle \mu - \mu^* , \nabla \Phi(\mu)\rangle + A \Phi^* +B \\
        & \leq A \| \mu - \mu^* \| \| \nabla \Phi(\mu) \|  + A \Phi^* +B 
    \end{aligned}
\end{equation*}
Solving this quadratic inequality gives
\begin{equation*}
\begin{aligned}
      \| \nabla \Phi(\mu)\|^2 & \leq (A \| \mu - \mu^*\| + \sqrt{A\Phi^* + B})^2 \\ 
      & \leq 2 (A^2 \| \mu - \mu^*\|^2 + A\Phi^* + B ).
\end{aligned}
\end{equation*}
}

\subsection{Resilience Analysis}
\proof[Proof of Lemma \ref{distanceofbregman}]{
First order condition of mirror step \eqref{mddyna} gives, for any $\mu \in \Delta$:
 $  \langle  \eta_t \ell^t  + \nabla \Psi (\mu^{t+1})-\nabla \Psi(\mu^{t}) ,  \mu^{t+1} - \mu \rangle \leq 0 $,
from which and Pythogarean identity:
    \begin{align*}
         & \quad D_{\Psi}(\mu, \mu^{t+1}) - D_{\Psi}(\mu^{t+1}, \mu^t) 
          \\
        & \leq \eta_t \langle \mu - \mu^{t+1},  \ell^t \rangle  - D_{\Psi} (\mu^{t+1}, \mu^t ) \\ 
        & = \eta_t \langle \mu - \mu^{t} , \ell^t \rangle  + \eta_t \langle \mu^{t} - \mu^{t+1} , \ell^t  \rangle  - D_{\Psi} (\mu^{t+1}, \mu^t ) \\ 
        & \leq \eta_t \langle \mu - \mu^{t} , \ell^t \rangle  + \eta_t \| \mu^{t} - \mu^{t+1} \| \| \ell^t \| - \frac{\sigma_{\Psi}}{2} \| \mu^{t} - \mu^{t+1}\|^2   \\
        & \leq \eta_t \langle \mu - \mu^t, \ell^t \rangle + \frac{2\eta_t^2}{\sigma_{\Psi}} (A \phi^t + B).
    \end{align*}
}

\proof[Proof of Lemma \ref{distanceofmus}]{
 For all $\mu \in \Delta$, since $\eta_t \leq \frac{1}{2A} \sigma_{\Psi}$, 
 \begin{equation*}
     \begin{aligned}
       & \quad D_{\Psi}( \mu, \mu^{t+1}) - D_{\Psi}(\mu , \mu^t ) \\ &\leq \eta_t (\phi(\mu, \omega^t) - \phi^t) + \frac{2\eta^2_t }{\sigma_{\Psi}} (A \phi^t + B) \\ 
      %
      & \leq \eta_t \phi(\mu, \omega^t) + \frac{B}{A} \eta_t .
     \end{aligned}
 \end{equation*}
 Plugging in $\mu = \mu^{t_0} \in \boldsymbol{\mu}^*$, we arrive at$ D_{\Psi}( \mu^*, \mu^{t+1}) - D_{\Psi}(\mu^* , \mu^t ) \leq \eta_t \phi^*(\omega^t ) + \frac{B}{A} \eta_t \leq \eta_t C_1$.

 Summing with respect to $t$, by strong convexity of $\Psi$ and taking infimum over $\Delta$, we get the result \eqref{dismus}. 
 Again taking $\mu = \mu^*$, with $\eta_t$ non-increasing we have 
 \begin{equation*}
     \begin{aligned}
       &  \frac{\eta_t}{2} \phi^t  \leq \eta_t \phi^*(\omega^t) +2 \frac{ B \eta_{t}^{2}}{\sigma_{\Psi}}+D_{\Psi}(\mu^*, \mu^{t})-D_{\Psi}(\mu^*, \mu^{t+1}) \\
  & \eta_t^2 \phi^t \leq 2 \eta^2_t C_1 + 2 \eta_t (D_{\Psi}(\mu^*, \mu^{t})-D_{\Psi}(\mu^*, \mu^{t+1})) \\ 
  & \leq 2C_1 \eta_t^2 + 2 \eta_t D_{\Psi}(\mu^*, \mu^{t})- 2 \eta_{t+1} D_{\Psi}(\mu^*, \mu^{t+1})
     \end{aligned}
 \end{equation*}
 Summing up the above, we arrive at $
\sum_{k=1}^{t} \eta_{k}^{2} \phi\left(\mu^{k}, \omega^{k}\right) \leq 2 C_{1} \sum_{k=1}^{t} \eta_{k}^{2} + 2\eta_1 a^{\dagger}$.
Note that $
    \phi^t \leq 2\phi^*(\omega^t) + 2 \frac{B}{A} + \frac{2}{\eta_t} (D_{\Psi}(\mu^*, \mu^{t})-  D_{\Psi}(\mu^*, \mu^{t+1}))$, and we obtain
 \begin{equation*}
 \begin{aligned}
     &  \sum_{k=1}^t \phi^k  \leq 2C_1 t +  \sum_{k=1}^t \frac{2}{\eta_k} (D_{\Psi}(\mu^*, \mu^{k})-  D_{\Psi}(\mu^*, \mu^{k+1})) \\
      & = 2C_1 t +  2\sum_{k=2}^t D_{\Psi}(\mu^*, \mu^{k}) (\frac{1}{\eta_k } - \frac{1}{\eta_{k-1}}) \\ & + 2\eta_1^{-1} D_{\Psi} (\mu^*, \mu^1) -  2\eta_t^{-1} D_{\Psi} (\mu^*, \mu^{t+1}) 
      \\& 
      \leq 2C_1 t + (2C_1 \sum_{k=1}^t \eta_k + a^{\dagger})\frac{1}{\eta_t}  + 2\eta_1^{-1} a^{\dagger} .
 \end{aligned}
 \end{equation*}
}
\vspace{-0.6cm}

\proof[Proof of Theorem \ref{distancemuwhp}] 
{
 We define the sequence $\xi_k := \eta_k \langle \mu^{t_0}  -  \mu^k, \ell^k - \E_{\omega^k}[\ell^k ] \rangle,\  t \in \mathbb{N}_+$. By lemma \ref{distanceofbregman}, plug in $\mu^{t_0} \in \boldsymbol{\mu}^*$,
 \begin{equation*}
     \begin{aligned}
     & D_{\Psi}(\mu^{t_0}, \mu^{t+1})-D_{\Psi}(\mu^{t_0}, \mu^{t})  \\ &
     \leq \eta_{t}\langle \mu^{t_0} - \mu^{t}, \ell^t\rangle  +\sigma_{\Psi}^{-1} \eta_{t}^{2} (A \phi^t+B) \\
     & = \xi_t + \eta_t\langle \mu^{t_0} - \mu^t , \mathbb{E}_{\omega^t} [\ell^t]  \rangle + \sigma_{\Psi}^{-1} \eta_{t}^{2} (A \phi^t+B)  \\
     & \leq  \xi_t + \eta_t( \Phi^* - \Phi^t)  + \sigma_{\Psi}^{-1} \eta_{t}^{2} (A \phi^t+B)
\end{aligned}
 \end{equation*}
It is easy to verify that $\mathbb{E}_{\omega^1, \ldots, \omega^k} [\xi_k]  = 0$, $(\xi_k)_k$ is thus a Martingale difference sequence. The conditional second moment of $\xi_k$ satisfies:$
      \mathbb{E}_k [|\xi_k|^2]   \leq \mathbb{E}_k [|\langle \mu^{t_0} -  \mu^k, \ell^k \rangle |^2] \leq \| \mu^{t_0} - \mu^k \|^2 \mathbb{E}_k [\|\ell^k \|^2 ] \leq  \| \mu^{t_0} - \mu^k \|^2  (A\Phi^k + B)$ .
Thus, the sum of conditional variances of $\xi_k$ is
\begin{equation*}
\begin{aligned}
       &\sum_{k=1}^t \mathbb{E}_k [|\xi_k - \E \xi_k |^2] = \sum_{k=1}^t \eta_k^2 \mathbb{E}_k [|\langle \mu^{t_0} -  \mu^k, \ell^k \rangle |^2] \\
   & \leq 2A \sigma_{\Psi}^{-1}\sum_{k=1}^t \eta_k ( \eta_1 a^{\dagger} + C_1  c_1 )  (\Phi^k - \Phi^*) \\
   &  \quad +   \sum_{k=1}^t \eta^2_k \| \mu^k - \mu^{t_0}\|^2  (A\Phi^* + B)  = : \sigma^2,
\end{aligned}
\end{equation*}
where we have let $c_1 := \sup _{k \in \mathbb{N}} \eta_k \sum_{j=1}^{k-1} \eta_j<\infty$.
From convexity of $\phi$, the magnitude  of the increments $c_2$ is as:
\begin{equation*}
\begin{aligned}
  &  \xi_k  - \mathbb{E}_k [\xi_k] = \eta_k \langle \mu^{t_0} - \mu^t, \ell^k \rangle + \eta_k \langle \mu^k - \mu^{t_0} , \mathbb{E}_k [\ell^k]\rangle \\ 
   & \leq \eta_k (\phi^*(\omega^k) - \phi^k) + \eta_k \| \mu^k - \mu^{t_0}\| \|\mathbb{E}_k[\ell^k] \| \\
& \leq \eta_k (\phi^*(\omega^k) - \phi^k)\\
&+ \frac{\eta_k}{2} ( (2\sigma_{\Psi}^{-1}C_1  \sum_{j=1}^{k-1}\eta_j +  2\sigma_{\Psi}^{-1}a^{\dagger}) + \|\mathbb{E}_k[\ell^k] \|^2) 
\\ 
& \leq \eta_k (\phi^*(\omega^k) - \phi^k) +\eta_k  ( \sigma_{\Psi}^{-1}( C_1  \sum_{j=1}^{k-1}\eta_j +  a^{\dagger})  \\ & \quad + A^2\|\mu^k - \mu^{t_0}\|^2 + A\Phi^* + B ) \\
& \leq \eta_1 (\phi^*  + A\Phi^* +B ) + \sigma_{\Psi}^{-1} (2 A^2 + 1)( C_1c_1  + \eta_1 a^{\dagger})
= : c_2
\end{aligned}
\end{equation*}
Using conditional Bernstein's inequality, for $\delta \in (0,1)$ let the constant $\rho := \min\{ 1, \frac{\sigma_{\Psi}}{ 2 A(\eta_1 a^{\dagger} + C_1 c_1)}  c_2\}$, one has with probability $1 - \delta$, $\sum_{k=1}^t \xi_k  - \mathbb{E}_k [\xi_k] = \sum_{k=1}^t \xi_k \leq  \frac{\rho \sigma^2}{c_2} + \frac{c_2 \log \frac{1}{\delta}}{\rho}$, 
we plug in the variance upper-estimate $\sigma^2$
and get $
\sum_{k=1}^t  \xi_k  \leq \sum_{k=1}^{t} \eta_{k}(\Phi^k-\Phi^{*}) \frac{ \sigma_{\Psi}\sum_{k=1}^t \eta^2_k \| \mu^k - \mu^*\|^2  (A\Phi^* + B)}{2 A(\eta_1 a^{\dagger} + C_1c_2) }  + \frac{c_2 \log (\frac{1}{\delta} ) }{\rho}$ .
Plugging in the inequalities in Lemma \ref{distanceofmus} and we have the $\Phi^k - \Phi^*$ term canceled due to $\rho$:
    $D_{\Psi}(\mu^{t_0}, \mu^{t+1}) \leq a^{\dagger} + \sigma_{\Psi}\frac{ \sum_{k=1}^t \eta^2_k \| \mu^k - \mu^{t_0}\|^2  (A\Phi^* + B)}{2A(\eta_1 a^{\dagger} + C_1c_2) }  + \frac{c_2 \log (\frac{1}{\delta} ) }{\rho} +  \sigma_{\Psi}^{-1} ((2A C_1 + B) \sum_{k=1}^t \eta_k^2 + 2 \eta_1 A a^{\dagger}) $,
with probability $1 - \delta$. By strong convexity of $\Psi$, the claim follows:
\begin{equation}\label{dismuwhp}
\begin{aligned}
    & \| \mu^{t+1} - \mu^{t_0} \|^2 \leq ( \frac{ 2}{\sigma_{\Psi}}  + \frac{4\eta_1 A}{\sigma_{\Psi}^{2}})a^{\dagger} +  \frac{ \sum_{k=1}^t \eta^2_k \| \mu^k - \mu^{t_0}\|^2 }{2(\frac{\eta_1}{C_1} a^{\dagger} + c_2) }  \\ & + \frac{2 c_2 \log (\frac{1}{\delta} ) }{\sigma_{\Psi}\rho} +  2 \sigma_{\Psi}^{-2} (2A C_1 + B) \sum_{k=1}^t \eta_k^2 .
\end{aligned}
\end{equation}

Define the event $\Omega_T$ as the following:
\begin{align*}
    &E_T   : =  \bigg\{ ( \omega_1, \ldots, \omega_T):  \forall t = 1, \ldots, T \text{ it satisfies } E_t \text{ where } \\ 
   & \| \mu^{t+1} - \mu^*\|^2  \leq ( \frac{ 2}{\sigma_{\Psi}}  + \frac{4\eta_1 A}{\sigma_{\Psi}^{2}})a^{\dagger} +  \frac{ \sum_{k=1}^t \eta^2_k \| \mu^k - \mu^{t_0}\|^2 }{2(\frac{\eta_1}{C_1} a^{\dagger} + c_2) }  \\  & + \frac{2 c_2 \log (\frac{T}{\delta} ) }{\sigma_{\Psi}\rho} +  \frac{4AC_1 +2 B }{\sigma_{\Psi}^2}  \sum_{k=1}^t \eta_k^2 \bigg\},
\end{align*}
by a union bound argument one has $\mathbb{P}\{E_T\} = 1 - \mathbb{P}(\bigcup_{t=1}^T E^{c, t}_T\} \geq 1 - \delta$. Since the series $\sum_{t=1}^{\infty} \eta_t^2$ converges, one can find $t_1 \in \mathbb{N}_+$ such that $ \sum_{k=t_1}^t \eta^2_k \| \mu^k - \mu^{t_0}\|^2   \leq  \frac{\eta_1}{C_1} a^{\dagger} + c_2$. Under $E_T$, one has that for all $t = 1, \ldots, T$, 
\begin{align*}
& \| \mu^{t+1} - \mu^{t_0}\|^2 -  \frac{2 c_2 \log (\frac{T}{\delta} ) }{\sigma_{\Psi}\rho} -\\ & ( \frac{ 2}{\sigma_{\Psi}}  + \frac{4\eta_1 A}{\sigma_{\Psi}^{2}})a^{\dagger}   -  \frac{4AC_1 +2 B }{\sigma_{\Psi}^2}  \sum_{k=1}^t \eta_k^2 \\
 \leq & \frac{\sum_{k=1}^{t_1} C_1 \sum_{j=1}^{k-1}\eta_j + a^{\dagger}}{\sigma_{\Psi}(\frac{\eta_1}{C_1} a^{\dagger} + c_2) } +   \frac{1}{2}\sup_{1 \leq \bar{k} \leq t}\| \mu^{\bar{k}} - \mu^{t_0}\|^2.
\end{align*}
Therefore, under the event $E_T$, we have:$
    \max_{1\leq t\leq T} \|\mu^t - \mu^{t_0}\|^2 \leq   \frac{4 c_2 \log (\frac{T}{\delta} ) }{\sigma_{\Psi}\rho} + ( \frac{ 4}{\sigma_{\Psi}}  + \frac{8\eta_1 A}{\sigma_{\Psi}^{2}})a^{\dagger}  +  \frac{8AC_1 +4 B }{\sigma_{\Psi}^2}  \sum_{k=1}^t \eta_k^2 + \frac{\sum_{k=1}^{t_1} 2 C_1 \sum_{j=1}^{k-1}\eta_j + a^{\dagger}}{\sigma_{\Psi}(\frac{\eta_1}{C_1} a^{\dagger} + c_2) }$, scaling the terms with $\log(\frac{T}{\delta})$, and replacing $\mu^{t_0}$ with $\inf_{\mu \in \boldsymbol{\mu}^*} $, we get the desired $C_2$.

}

\proof[Proof of Prop. \ref{resilienceprop}]{
Now going back to the offset term $\sum_{k=1}^t \Phi^k - \Phi^*$. By lemma \ref{onestepinequal}, we have
\begin{align*}
& \quad  D_{\Psi}(\mu^{t_0}, \mu^{t+1})-D_{\Psi}(\mu^{t_0}, \mu^{t}) - 2 \frac{\eta_{t}^{2}}{\sigma_{\Psi}}\left(A \phi^{t}+B\right)\\ & \leq \eta_{t}\left\langle\mu^{t_0}-\mu^{t}, \ell^{t} - \mathbb{E}_{\omega^t} [\ell^t]\right\rangle + \eta_{t}\left\langle\mu^{t_0}-\mu^{t}, \mathbb{E}_{\omega^t} [\ell^t]\right\rangle \\
& \leq \xi_t + \eta_t (\Phi^*- \Phi^t) ,
\end{align*}
where the last inequality is by convexity, taking summation over $k = 1, \ldots, t$, $\sum_{k=1}^t \eta_k (\Phi^k - \Phi^*)  \leq a^{\dagger} + \sum_{k=1}^t \xi_k + 2 \frac{\eta_{k}^{2}}{\sigma_{\Psi}}\left(A \phi^{k}+B\right) \leq (1 + \frac{2A \eta_1}{\sigma_{\Psi}})a^{\dagger}  + \sum_{k=1}^t \xi_k + \frac{2(AC_1 + B)}{\sigma_{\Psi}} \sum_{k=1}^t \eta_k^2 $.
Let $ \xi^{\prime}_t := \eta_t \langle \mu^{t_0}  - \mu^t,  \ell^t - \mathbb{E}_{\omega^t} [\ell^t]\rangle \mathds{1}_{ \{ \|\mu^t - \mu^{t_0}\|^2 \leq C_2 \log(\frac{2T}{\delta})\} }$, 
we can estimate its magnitude $b(\mu)$ by the following manipulation, under the event $\{\|\mu^t - \mu^{t_0}\|^2 \leq C_2 \log(\frac{2T}{\delta})\}$, by Cauchy Schwarz and triangular inequality, together with Lemma \ref{boundphik},
\begin{align*}
    | \xi^{\prime}_t  | & \leq \eta_t [ \|\mu^t - \mu^{t_0}\|^2 + \| \ell^t\|^2 + \|\mathbb{E}_{\omega^t}[\ell^t]\|^2 ] \\
    & \leq \eta_t [ (4A^2 + 1) \|\mu^t - \mu^{t_0}\|^2 + 2 A (\phi^* + \Phi^*) + 4B ] \\
    & \leq \eta_t [ (4A^2 + 1) C_2 \log(\frac{2T}{\delta}) + 4A C_1 ] \leq  c_3 \eta_t \log(\frac{2T}{\delta}).
  \end{align*}
Therefore, with probability $1 - \frac{\delta}{2}$ one can find a $E^{\prime}_T$ such that by Lemma \ref{concentrationtech} i., the following inequality holds:$
    \sum_{k=1}^T \xi^{\prime}_k \leq c_3 \log(\frac{2T}{\delta})(2\sum_{k=1}^T \eta_k^2 \log\frac{2}{\delta})^{\frac{1}{2}}  \leq  c_3 \log ^{\frac{3}{2}} \frac{2 T}{\delta}(2 \sum_{t=1}^{T} \eta_{t}^{2})^{\frac{1}{2}}$.
Let $E_T$ be such that $\max_{ 1 \leq t \leq T} \|\mu^t  - \mu^{t_0} \|^2 \leq C_2 \log(\frac{2T}{ \delta})$. With a union bound argument, $\mathbb{P}\{E_T \bigcap E^{\prime}_T\} = 1 - \mathbb{P}\{E^c_T \bigcup E^{\prime c}_T\} \geq 1 - \delta$, in which case, 
\begin{align*}
    & \sum_{k=1}^t \eta_k (\Phi^k - \Phi^*)  \leq (1 + \frac{2A \eta_1}{\sigma_{\Psi}})a^{\dagger}\\ & +  c_3 \log ^{\frac{3}{2}} \frac{2 T}{\delta}(2 \sum_{t=1}^{T} \eta_{t}^{2})^{\frac{1}{2}}   + \frac{2(AC_1 + B)}{\sigma_{\Psi}} \sum_{k=1}^t \eta_k^2  \\
   &   \leq \bigg( (1 + \frac{2A \eta_1}{\sigma_{\Psi}})a^{\dagger} + ((4A^2+ 1)C_2 + 4AC_1)(2\sum_{t=1}^{\infty}\eta_t^2)^{\frac{1}{2}}  \\ &
     + \frac{2(AC_1 + B)}{\sigma_{\Psi}} (\sum_{k=1}^{\infty} \eta_k^2) \bigg) \log^{\frac{3}{2}}(\frac{2T}{\delta}) 
     = : C_3 (a^{\dagger})\log^{\frac{3}{2}} (\frac{2T}{\delta}). 
\end{align*}
Using the convexity of $\Phi$, we arrive at the result.
}

\proof[Sketched Proof of Proposition \ref{examprop}]{
Let $\Psi(\mu) = \sum_{w \in \mathcal{W}} \Psi_w(\mu)$ where $\Psi_w(\mu) = \sum_{p \in \mathcal{P}_w} \mu_p \log \mu_p - \mu_p$. It is obvious that the function $\Psi_w(\mu)$ is $\frac{1}{m_w}$-strongly convex on $\{ (\mu_p)_{p \in \mathcal{P}_w} : \sum_{p \in \mathcal{P}_w}\mu_p = m_w\}$, for sub-gradient $s \in \partial \Psi (\mu^{\prime} )$, 
 \begin{align*}
    \Psi(\mu) - \Psi(\mu^{\prime}) & \geq \langle s, \mu - \mu^{\prime}\rangle + \sum_{w \in \mathcal{W}} \frac{1}{2 m_w} \sum_{p \in \mathcal{P}_w}( \mu_p - \mu^{\prime}_p )^2 \\
     & \geq  \langle s, \mu - \mu^{\prime}\rangle + \frac{1}{2 M} \|\mu - \mu^{\prime}\|^2
 \end{align*}

 The attack is the KL divergence under the choice of $\Psi$,$
     a^{\dagger} = D_{\Psi}( \mu^{t_0} , \mu^{\dagger})  = \sum_{p \in \mathcal{P}} \mu^{t_0}_p \log(\frac{\mu^{t_0}_p}{\mu^{\dagger}_p})$. 
 Let $ \mathrm{supp}(\mu^{t_0}) \subseteq \mathrm{supp}(\mu^{\dagger}) $ such that $a^{\dagger}$ is finite, let $\gamma_w = \min_{p \in\mathcal{P}_w, \mu_p > 0} \mu^{\dagger}_p $. 
 by Hölder's inequality (the lower bound) and reverse Pinsker's inequality (the upper bound),

 \begin{equation*}
    \sum_{p \in \mathcal{P}} \mu^{t_0}_p \log\left( \frac{|\mathcal{P}|\mu_p^{t_0}}{\bar{M}}\right)  \leq a^{\dagger} \leq   \frac{\|\mu^{t_0} - \mu^{\dagger} \|^2_1}{ \min_{w \in \mathcal{W}}\gamma_w \ln2}, 
 \end{equation*}
 the first equality holds when $\mu^{\dagger}$ is such that every path has equally distributed flow. 
 By triangular inequality, the $a^{\dagger}$ is bounded by $\frac{4\bar{M}^2}{\min_{w\in\mathcal{W}}\gamma_w \ln 2}$. Plugging in $a^{\dagger}$ into Proposition \ref{resilienceprop} yields the results. 
}


\end{document}